\documentclass[a4paper,10pt]{article}
\usepackage[utf8]{inputenc}
\usepackage{amsmath,amsthm,amssymb}
\usepackage{enumerate}
\usepackage{hyperref}
\usepackage{doi}
\usepackage{graphicx}
\usepackage{subfigure}
\usepackage{xcolor}
\usepackage{cite}
\usepackage{dsfont} \usepackage{stmaryrd} \usepackage{authblk} \usepackage{tabularx}
\usepackage{subcaption}
\usepackage{geometry}
\usepackage{tikz}
\usepackage{color}
\usepackage{xcolor}
\usetikzlibrary{arrows.meta}
\usetikzlibrary{decorations.pathreplacing}
\usepackage[english]{babel}

\date{}

\theoremstyle{definition}

\theoremstyle{definition}

\newcommand{\RR}{\mathds{R}}

\newcommand{\ddt}[1]{\frac{d #1}{dt}}

\newcommand{\clspeed}{V_\Gamma}
\newcommand{\sigmawet}{\sigma_{\text{w}}}
 
\DeclareMathOperator{\thetam}{\theta_m}  

\title{On the problem of optimal fluid transport in capillaries}
\author[1]{Mathis~Fricke\thanks{fricke@mma.tu-darmstadt.de}}
\author[2]{Clara~Bernklau}
\author[2]{Elisabeth~Diehl}
\author[3]{Joël~De~Coninck}
\author[2]{Stefan~Ulbrich}
\author[1]{Dieter~Bothe}

\affil[1]{Mathematical Modeling and Analysis Group, TU Darmstadt, Germany}
\affil[2]{Fachbereich Mathematik, TU Darmstadt}
\affil[3]{Transfers, Interfaces and Processes, Université libre de Bruxelles, Belgium}

\usepackage{color}

\begin{document}
\maketitle

\begin{abstract}
 In this note, we revisit the problem of the pressure-driven transport of a meniscus through a narrow cylindrical capillary or pore. This generic process finds many applications in science and technology. As it is known that Direct Numerical Simulations of moving contact line problems are highly demanding in terms of computational costs, simplified models in the form of ordinary differential equations offer an interesting alternative to perform a mathematical optimization of the flow. Blake and De~Coninck studied the pressure-driven transport of a meniscus and identified two major competing mechanisms. While a hydrophilic surface is favorable to enhance the spontaneous imbibition into the pore, the friction is known to be significantly reduced on a hydrophobic surface. Blake and De~Coninck showed that, depending on the applied pressure difference, there exists an optimal wettability that minimizes the time required to move the meniscus over a certain distance. We revisit this problem and derive analytical solutions in the limiting cases of negligible inertia and negligible contact line friction.
\end{abstract}

\section{Introduction}
Understanding the propagation of a liquid within a pore network due to pressure gradients presents a multifaceted challenge rooted in the complex dynamics of fluid flow through porous media. Prominent instances encompass the extraction of petroleum \cite{Mason2013}, geological sequestration of CO$_2$ \cite{Bachu2000} and the remediation of groundwater contaminants \cite{AlHashimi2021} within the realm of engineering. Additionally, precise drug delivery and the dynamics of gas–liquid flow are pivotal in the pharmaceutical domain \cite{Maged2022}. Meanwhile, in the industrial sphere, microfluidic-based chemical and physical processes hold significance \cite{Cheng2023}. The dynamics of fluid displacement in these contexts stand as paramount concerns. At the heart of this challenge lies the intricate interplay of various factors governing fluid behavior across both micro and macro scales.\\
\\
One primary complexity lies in the irregular geometry of pores found in porous media, which is essential for predicting how pressure gradients will affect the movement of liquid within the pore network. Capillary forces further complicate the picture, exerting significant influence on fluid behavior. Surface tension, contact angles, and pore size distribution all play crucial roles in determining whether and how fast a liquid column will rise or fall within the pores, adding another layer of complexity to the characterization process. Moreover, the heterogeneity inherent in porous media introduces non-uniformities in fluid flow patterns. Variations in pore size, shape, and connectivity lead to diverse flow behaviors, making it challenging to predict the propagation of a liquid column accurately.
Accounting for these effects is crucial for accurately characterizing liquid column propagation under pressure gradients.
Additionally, fluid flow in porous media occurs across a wide range of time and length scales, from pore-scale phenomena like viscous fingering to macroscopic phenomena like Darcy flow. Bridging the gap between these scales is essential for gaining a comprehensive understanding of liquid column propagation.
While theoretical models offer insights into fluid behavior, experimental validation is indispensable. However, conducting experiments in realistic porous media environments presents challenges such as scale-up, accessibility, and measurement limitations. Overcoming these challenges requires a multidisciplinary approach that combines theoretical modeling, experimental techniques, and computational simulations to unravel the complexities of fluid flow in porous media under pressure gradients.
As a first step, we revisit here the propagation of a liquid into a single pore.
Our goal is to derive (for a simplified model) the exact solution for the propagation of a liquid column into a single pore.
Firstly, it would provide insights into the fundamental processes governing fluid flow at the pore scale. Understanding these processes, such as capillary action and interfacial interactions, is important for a wide range of applications, including groundwater hydrology, oil recovery, and filtration systems.
Understanding the behavior of fluid flow at the pore scale serves as a fundamental basis for comprehending the dynamics of fluid movement throughout the entire porous media network. By delving into the intricacies of how fluids behave within individual pores, we may hope to extrapolate this knowledge to predict how fluid flow interacts and propagates across the entire network.
Moreover, the exact solution acts as a crucial reference point for validating and refining multiscale models that aim to bridge the gap between pore-scale phenomena and macroscopic behavior. Comparing the predictions of these models with the analytical solution at the single pore level helps ensure their accuracy and reliability when applied to more complex pore networks.
Additionally, the insights gained from the exact solution inform parameter estimation and calibration processes in complex pore network models. By understanding how parameters such as pore geometry and fluid properties influence fluid behavior at the single pore level, researchers can effectively tune these parameters to accurately represent experimental observations in more complex systems.\\
\\
Let us introduce the variables of the problem. The pore itself is considered a perfect cylinder with radius $R$. The liquid is characterized by its density $\rho$, its viscosity $\eta$ and surface tension $\sigma$. The affinity between the liquid and the solid is quantified by the contact angle at equilibrium $\theta_0$. The imposed pressure difference is denoted $\Delta P \geq 0$. We denote by $h(t)$ the imbibition height (or length) at time $t$. An optimization of the transport process is interesting to, e.g., maximize the imbibition speed or to minimize the energy consumption. Blake and De~Coninck \cite{Blake2004} showed that imbibition speed can be maximized with respect to the wettability of the surface. This can be achieved in practice by a suitable surface coating. In this note, we derive an analytical expression for the optimal wettability, resulting from the model by Blake and De~Coninck.\\

\definecolor{MyBlue}{rgb}{0.9,1.5,1.5}
\begin{figure}[hb]
\centering{
\begin{tikzpicture}[>={Triangle[length=0pt 9,width=0pt 5]}]
\begin{scope}[rotate = -90]
\draw [color=MyBlue, fill=MyBlue] (2,-0.2)  -- (2,4.5) to [out=-45, in=-135] (4,4.5) -- (4,-0.2) --(2,-0.2);

\draw[ line width=2pt] (2,-0.2) -- (2, 5.5);
\draw[ line width=2pt] (4,-0.2) -- (4, 5.5);

    \draw[-, thick] (2,4.5) to[out=-45, in=-135] (4,4.5);
    
  \draw[<->] (2,1.7) to (4, 1.7);
  \draw[-] (3,4.1) to (3, 4.3);
    \draw[-] (3.1,4.1) to (5, 4.1);
      \draw[<->] (5,4.1) to (5, 1);

 \draw[->] (1.8,1.8) to (1.8, 3);
  \draw[->] (1.8,1.8) to (1.8, 3);
  \draw[dashed] (2,1) -- (4,1);
    \draw[-] (4,1) -- (5,1);
    \draw[-] (2,4.5) -- (2.8,3.8);
     \draw[-] (2,3.8) to[out=-60, in=-100] (2.58,4);

\node (C) at (3,3.2)  {\scriptsize $\rho, \eta$};  
\node (D) at (3,4.8)  {\scriptsize $\sigma>0$};  
\node (A) at (3,2)  {\scriptsize $2R$};
\node (E) at (5.3,2.5)  {\scriptsize $\Delta h$};
\node (F) at (5.3,1)  {\scriptsize $h_0$};
\node (G) at (2.2, 4)  {\scriptsize $\theta$};
\node (B) at (1.5,2.4)  {\scriptsize $h$, $\dot{h}$, $\Delta P$};
  
\end{scope}
\end{tikzpicture}}
    \caption{Sketch of the problem for a fluid traveling from $h(0)=h_0$ to $h_0+\Delta h$.}
\end{figure}
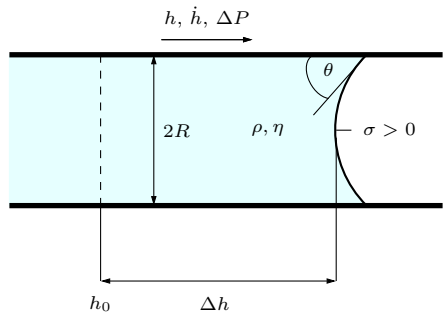

Following \cite{Blake2004}, the single pore system can be modeled reasonably well by an ordinary differential equation which approximates the forces acting on the liquid column. It is based on the work by Bosanquet \cite{Bosanquet1923}, and later Martic et al.\ \cite{Martic2002}. The model reads as
\begin{align}\label{eqn:blake2004-model}
\pi R^2 \Delta P + 2 \pi R \sigma \cos \theta_0 - 2 \pi R \zeta \dot{h} = 8 \pi \eta h \dot{h} + \ddt{} \left(\pi R^2 \rho h \dot{h} \right).
\end{align}
Here $\Delta P > 0$ denotes an externally applied pressure difference that drives the flow. The capillary force $2 \pi R \sigma \cos \theta_0 = -2 \pi R \sigmawet$ sucks the liquid into the capillary if the surface is hydrophilic (i.e., if $\theta_0 < \pi/2$) or pushes it out of the capillary if the surface is hydrophobic (i.e., if $\theta_0 > \pi/2$). The contact line friction force $2 \pi R \zeta \dot{h}$ is one of the major dissipative mechanisms in the system and is linked to the dynamics of the microscopic contact angle $\thetam$ according to
\begin{align}\label{eqn:linear-cl-friction}
- \zeta \clspeed = \sigma (\cos \thetam - \cos \theta_0).
\end{align}
Here $\clspeed$ denotes the speed of the contact line. In our case, it can be identified with the rate-of-change of the imbibition height, i.e.
\[ \clspeed = \dot{h}. \]
The viscous force $8 \pi \eta h \dot{h}$ arises from the viscous friction due to the Hagen-Poiseuille flow inside the liquid column sufficiently away from the interface. It is the second dissipative mechanism present in the model \eqref{eqn:blake2004-model}. Finally, the last term in \eqref{eqn:blake2004-model} describes the rate-of-change of the total momentum inside the liquid column. We will refer to it as the inertial term in the following. It is, however, important to note that these expressions are all \emph{approximations} to the forces in a full continuum mechanical description of the system (see \cite{Fricke2023,Quere1997,Quere1999,Zhmud2000} for more details on the modeling). A more accurate description can be achieved by using CFD simulations based on the two-phase Navier Stokes equations (see, e.g., \cite{Sprittles2011b,Yamamoto2014,Gruending2020b,Keeler2022}). However, these simulations have much higher computational costs due to the multiscale nature of the moving contact line problem \cite{PreprintFricke2023a}. Therefore, simplified models like \eqref{eqn:blake2004-model} offer an interesting alternative.\\
\\
Recently, it has been shown that the friction mechanism described by the parameter $\zeta$ is related to the fluctuations of the contact line \cite{Fernandez-Toledano2019,FernandezToledano2020}. The importance of the fluctuations is related to the size of the system, i.e.\ the value of the radius $R$ in the present case. In small systems, such as those consisting of a small number of molecules, fluctuations can indeed have a significant impact due to the relatively small number of constituent particles. Because there are fewer particles to average out the effects of random motions, the system's properties can fluctuate more dramatically. These fluctuations can influence various thermodynamic properties such as, here, the dissipation of energy at the contact line. In large systems with a large number of molecules, fluctuations tend to average out due to the law of large numbers. The behavior of individual particles becomes less significant compared to the overall behavior of the system, leading to smoother, more predictable properties. Eq~\eqref{eqn:blake2004-model} is thus particularly well suited to describe how a liquid column is propagating into a micro pore due to a gradient of pressure $\Delta P$.

\subsection{The Lucas-Washburn equation}
In the seminal work by Lucas and Washburn \cite{Lucas1918,Washburn1921}, a simplified version of \eqref{eqn:blake2004-model} is studied, where the viscous friction due to the Poiseuille flow is balancing the capillary driving force. In this case, the model reads as
\begin{align}\label{eqn:washburn-model}
2 \pi R \sigma \cos \theta_0 = 8 \pi \eta h \dot{h}.
\end{align}
We transform \eqref{eqn:washburn-model} into non-dimensional form by choosing the radius $R$ as the length scale and (yet unspecified) $\tau$ as a time-scale. This yields the problem
\[ \tau \, \frac{\sigma \cos \theta_0}{4 \eta R} = H(s) H'(s). \]
This suggests choosing the time-scale $\tau = 4 \eta R/\sigma$ leading to the ODE
\begin{align}
H(s) H'(s) = \cos \theta_0.
\end{align}
The goal is now to compute the time that it takes the fluid front to travel from $H_0$ to $H_0 + \Delta H$. In the following, we apply a phase space approach (see, e.g., \cite{Hartmann2021a,Walter1998}) to solve the ODE problem. This approach is convenient here because it can also be applied for more general situations (see below). We write the imbibition speed $V = H'$ as a function of the penetration height $H$, i.e.
\begin{align}
H'(s) = \frac{\cos \theta_0}{H(s)} \quad \Rightarrow \quad V(H) = \frac{\cos \theta_0}{H}.
\end{align}
Then, the elapsed (non-dimensional) time to move the meniscus from $H_0$ to $H_0 + \Delta H$ is given as the integral of the inverse speed, i.e.\
\begin{align}
\Delta s = \int_{H_0}^{H_0+\Delta H} \frac{dH}{V(H)} = \frac{1}{\cos \theta_0} \int_{H_0}^{H_0 + \Delta H} H \, dH = \frac{1}{2 \cos \theta} ((H_0 + \Delta H)^2 - H_0^2).
\end{align}
Converting back this relation into physical units leads to
\begin{align}
\Delta t = \frac{2 \eta}{R \sigma \cos \theta_0} \Delta h (\Delta h + 2 h_0).
\end{align}

\subsection{Non-dimensional form of the governing equation}
Using the length and time scales from the Lucas-Washburn equation (i.e., $R$ and $\tau=4\eta R/\sigma$), we may write equation \eqref{eqn:blake2004-model} in non-dimensional form as
\begin{align}\label{eqn:blake2004-model-nondim}
\boxed{\Delta \tilde{P} + \cos \theta_0 = H H' + \frac{1}{32 \, \text{Oh}^2} (H H')' + \frac{1}{4} \tilde{\zeta} H',}
\end{align}
where
\[ \text{Oh} = \frac{\eta}{\sqrt{R \rho \sigma}}, \quad \tilde{\zeta} = \zeta/\eta, \quad \Delta \tilde{P} = \Delta P/(2\sigma/R). \]
Notice that the Ohnesorge number $\text{Oh}$ appears as a non-dimensional parameter. Below, we will derive analytical solutions of this problem in the limiting cases of negligible inertial effects and negligible friction $\tilde\zeta$.

\section{Analytical solution and optimal transport in the absence of inertial effects}
For large values of the Ohnesorge number, the inertial term in \eqref{eqn:blake2004-model-nondim} may be neglected, leading to
\begin{align}\label{eqn:nondim-martic-model-without-inertia}
\Delta \tilde{P} + \cos \theta_0 = H H' + \frac{1}{4} \tilde{\zeta} H'.
\end{align}
We rewrite \eqref{eqn:nondim-martic-model-without-inertia} in the phase space formulation as
\begin{align}\label{eqn:governing-equation-without-inertia}
H'(s) = \frac{\Delta \tilde{P} + \cos \theta_0}{H + \tilde\zeta/4} = V(H(s)).
\end{align}
Assuming
\[ \alpha := \Delta \tilde{P} + \cos \theta_0 > 0, \]
we can directly compute the required time to move the meniscus using the relation
\begin{align*}
\Delta s = \int_{H_0}^{H_0 + \Delta H} \frac{dH}{V(H)} = \frac{1}{\Delta \tilde{P}+\cos\theta_0} \int_{H_0}^{H_0+\Delta H} \left(H+\frac{1}{4} \tilde{\zeta} \right) dH = \frac{ \Delta H \left( H_0 + \frac{\Delta H}{2} + \frac{1}{4} \tilde{\zeta} \right)}{\Delta \tilde{P}+\cos\theta_0}.
\end{align*}
Converting back the expression to physical units leads to the expression
\begin{align}\label{eqn:imbibition_time_martic_type}
\Delta t = \frac{4 \eta \Delta h \left( \frac{h_0}{R} + \frac{\Delta h}{2R} + \frac{1}{4} \frac{\zeta}{\eta} \right)}{\sigma \cos \theta_0 + R \Delta P/2}.
\end{align}

\begin{figure}[ht]
	\begin{minipage}[c]{.48\linewidth}
	\includegraphics[width=\textwidth]{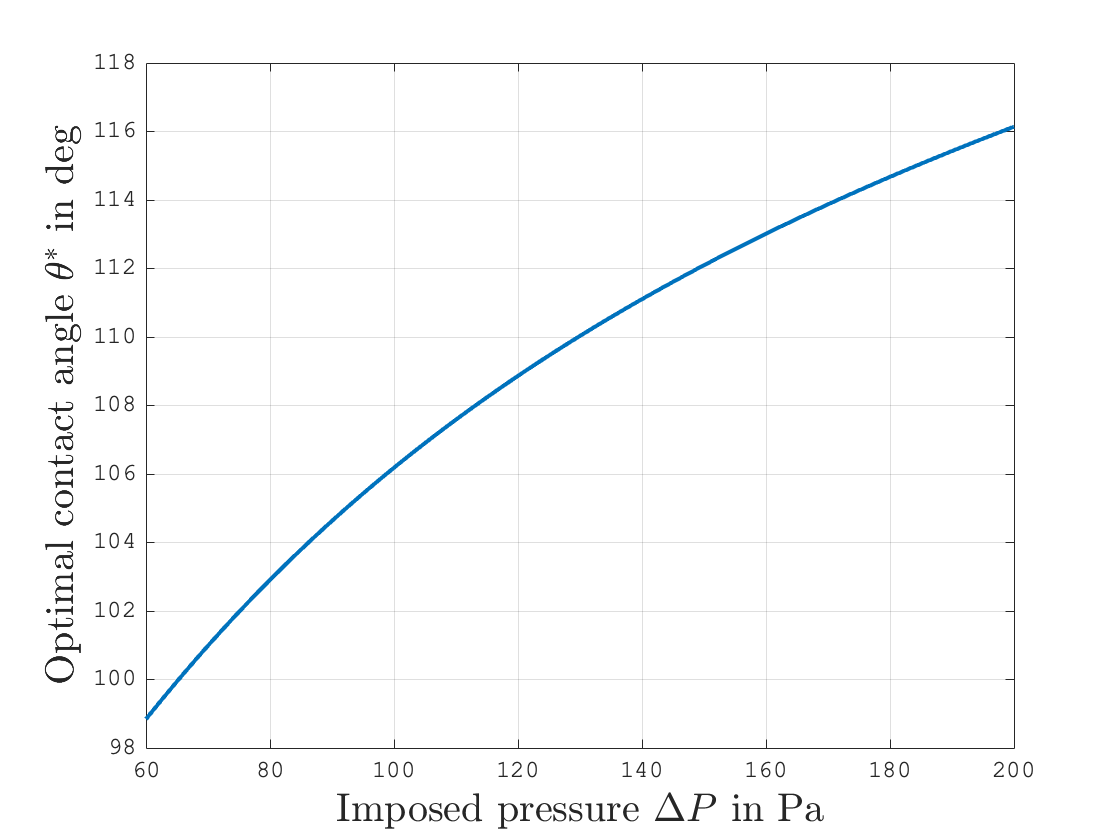}
  \caption{Optimal contact angle in dependence of the imposed pressure for water, see Table \ref{table1}. We choose $R=1$ mm, $h_0=1$ cm, $\Delta h=10$ cm, $\Delta P=150$ Pa and $T=297.15$ K.}
	\label{fig:pressure_angle}
	\end{minipage}
	\hfill
	\begin{minipage}[c]{.48\linewidth}
	\includegraphics[width=\textwidth]{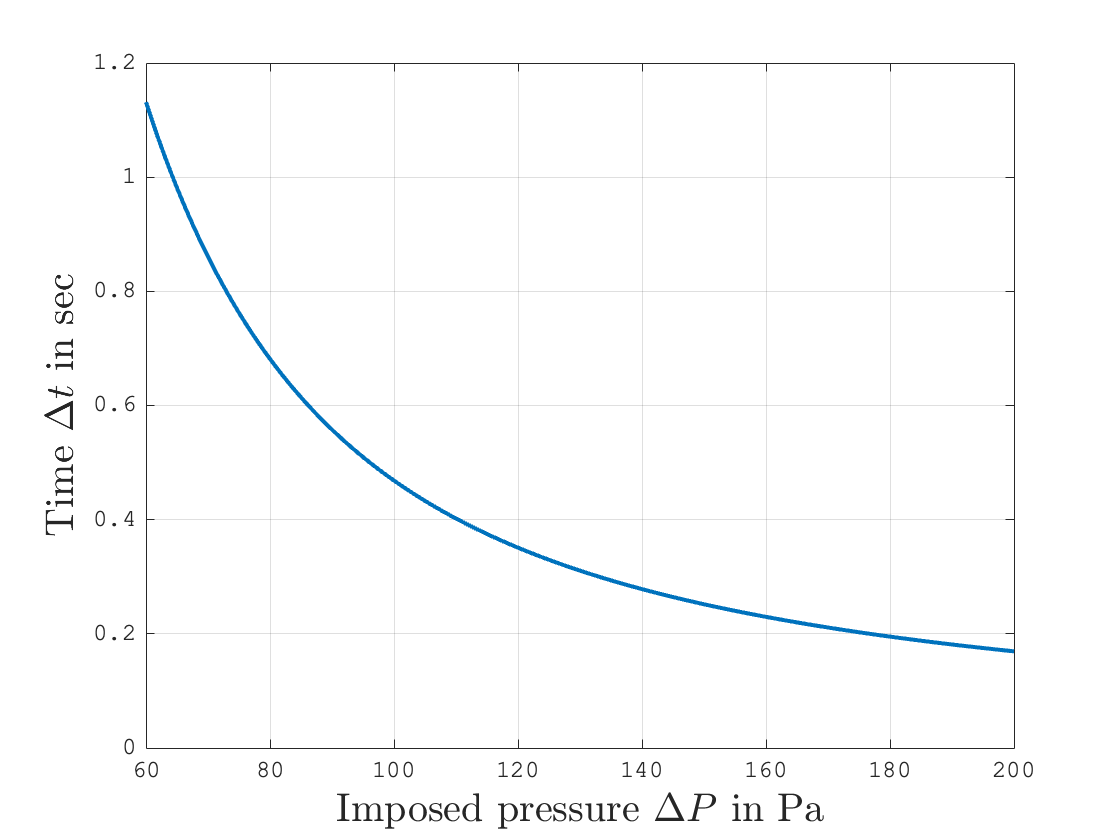}
  \caption{Optimized time in dependence of the imposed pressure for water. The parameters and setup are the same as for Figure \ref{fig:pressure_angle}.}
	\label{fig:pressure_time}
	\end{minipage}
\end{figure}

\paragraph{Optimal wettability:} In \cite{Blake2004}, it has been shown that there exists an \emph{optimal wettability} for the transport of liquid within a pore. This is a consequence of the competition of two important influencing factors in \eqref{eqn:imbibition_time_martic_type}, namely
\begin{enumerate}[(i)]
\item the surface should be hydrophilic, i.e.\ $\sigma \cos \theta_0$ should be large to increase the driving force,
\item but on the other hand, the surface should be hydrophobic to decrease the friction coefficient. Blake and De~Coninck use the following model for the contact line friction coefficient (see \cite{Blake2004,Blake2015})
\begin{align}\label{eqn:zeta_theory_mkt}
\zeta = \eta \frac{V_l}{\lambda^3} \exp\left( \frac{\sigma(1+\cos \theta_0)\lambda^2}{k_B T} \right).
\end{align}
\end{enumerate}
We will now derive the optimal wettability from the expressions \eqref{eqn:imbibition_time_martic_type} and \eqref{eqn:zeta_theory_mkt}. Combining \eqref{eqn:zeta_theory_mkt} and \eqref{eqn:imbibition_time_martic_type} leads to \begin{align*}
\Delta t = \frac{4 \eta \Delta h \left( \frac{h_0}{R} + \frac{\Delta h}{2R} + \frac{1}{4 \eta} \eta \frac{V_l}{\lambda^3} \exp\left( \frac{\sigma(1+\cos \theta_0)\lambda^2}{k_B T} \right) \right)}{\sigma \cos \theta_0 + R \Delta P/2}.
\end{align*}
From \cite{Blake2004}, we know that $V_l$ is proportional to $\lambda^3$, where we assume this constant is equal to one. We thus obtain
\begin{align}\label{eqn:delta_t}
\Delta t = \frac{\eta \Delta h \left( \frac{4 h_0}{R} + \frac{2 \Delta h}{R} + \exp\left( \frac{\sigma(1+\cos \theta_0)\lambda^2}{k_B T} \right) \right)}{\sigma \cos \theta_0 + R \Delta P/2}.
\end{align}
Our goal is now to find a formulation for a critical point of the required time with respect to $\cos \theta_0$. Since \eqref{eqn:delta_t} is differentiable, we can apply the necessary first order optimality condition to calculate a local minimum $\bar{x}$ by setting the derivative equal to zero. Doing so, we substitute some quantities in \eqref{eqn:delta_t} for a clearer presentation, viz.\ let
\begin{align*}
a := \frac{4 h_0+2 \Delta h}{R}, \quad b := \frac{R \Delta P}{2\sigma}, \quad c := \frac{\sigma \lambda^2}{k_B T}.
\end{align*}
Furthermore, we write $x=\cos \theta_0$. This leads to
\begin{align*}
\Delta t = \frac{\eta \Delta h \left(a + \exp\left(c (1+x)\right) \right)}{\sigma (x + b)}.
\end{align*}
Now we obtain for the derivative of $\Delta t$ with respect to $x$ the identity
\begin{align*}
\Delta t' &= \frac{\eta \Delta h \sigma \left(\exp\left(c (1+x)\right) (c (x + b) - 1) - a \right) }{\left( \sigma (x + b)\right)^2}.
\end{align*}
The above expression has to be equal to zero to fulfill the required optimality condition. The right-hand side can only be equal to zero if the numerator is equal to zero, so necessarily
\begin{align*}
\exp\left(c (1+x)\right) (c x + b c - 1) = a.
\end{align*} 
Such a type of equation can be solved with the Lambert $W$ function to the result
\begin{align*}
x = \frac{1}{c} \left(W(a \exp\left((b-1)c - 1 \right)) - b c + 1\right).
\end{align*} 
Hence, a critical point of the required time appears at
\begin{align}\label{eqn:optimal_contact_angle_martic}
\boxed{\cos \theta_0^\ast = \frac{k_B T}{\sigma \lambda^2} \left( W\left( \frac{4 h_0+2 \Delta h}{R} \exp\left(\left(\frac{R \Delta P}{2\sigma} - 1 \right) \frac{\sigma \lambda^2}{k_B T} - 1\right) \right) - \frac{R \Delta P \lambda^2}{2k_B T} + 1 \right).}
\end{align}
From equation~\eqref{eqn:optimal_contact_angle_martic}, we can immediately infer some physical properties of the optimal solution; see Figures~\ref{fig:pressure_angle} and \ref{fig:pressure_time}. In particular, we see that the optimal contact angle is monotonically increasing with the imposed pressure $\Delta P$. This is because a low value of the friction coefficient defined in the model~\eqref{eqn:zeta_theory_mkt} is preferred for a high external forcing. On the other hand, the required time in the optimal state is monotonically decreasing with the imposed pressure (see Fig.~\ref{fig:pressure_time}).

\section{Inertial effects in the absence of contact line friction}
Another limiting case that allows for an analytical solution is the one with vanishing friction $\tilde\zeta$. In this case, we obtain the problem
\begin{align}\label{eqn:vanishing-zeta-model}
\alpha = HH' + \varepsilon (HH')',
\end{align}
where $\alpha = \Delta \tilde{P} + \cos \theta_0$ and $\varepsilon = 1/(32 \, \text{Oh}^2)$ are positive constants. To solve \eqref{eqn:vanishing-zeta-model}, we substitute
\[ z(s) = H(s)H'(s) = \frac{1}{2} \frac{\text{d}}{\text{ds}} H(s)^2. \]
Obviously, $H(s)$ can be inferred from the integral of $z(s)$. The equation for $z$ simply reads as
\begin{align}
\alpha = z + \varepsilon z'
\end{align}
with the general solution
\[ z(s) = \alpha + (z(0)-\alpha) \, e^{-s/\varepsilon}, \quad z(0) \in \RR. \]
We hence obtain
\begin{equation}
\begin{aligned}
H(s)^2 &= H(0)^2 + 2 \int_0^s z(y) \, dy \\
&= H(0)^2 + 2 \alpha s + 2 \varepsilon (H(0)H'(0)-\alpha) (1-e^{-s/\varepsilon}).
\end{aligned}
\end{equation}

\section{Optimization procedure}
    Since the goal of this paper is to minimize the wetting time in a horizontal pore and thus to maximize the speed of wetting for a liquid that has already reached a length of $h_0$ to wet another $\Delta h$, the optimization problem for the contact angle $\theta$ and the analytical solution of $\Delta t$ has the following form:
    \begin{align*}
        \min_{\theta} \Delta t \quad \text{s.t.}  \quad \Tilde{\zeta} = \eta \exp \Biggl(\frac{\sigma(1+ \cos( \theta) \lambda^2}{k_B T} \Biggr), \\
         \Delta s = \frac{\Delta H (H_0 + \frac{\Delta H}{2}+\frac{\Tilde{\zeta}}{4})}{\Delta \Tilde{P} + \cos(\theta)}, \\
        \Delta t = \tau \Delta s, \\
         \quad  b_l \leq \theta \leq b_u, \\
    \end{align*}
    where $H_0 >0$ is the initial length that has already been reached by the liquid, $b_l \in \mathbb{R}$ is the lower bound of $\theta$ and $b_u \in \mathbb{R}$ the upper bound. We choose $b_l = 0^{\circ}$, $b_u=180^{\circ}$ for all our investigations.  \\ 
Moreover, instead of using the analytical solution, we also solve the following equivalent ODE-constrained optimization problem by using the model \eqref{eqn:governing-equation-without-inertia}:
    \begin{align*}
        \min_{\theta} \Delta t \quad \text{s.t.}  \quad \Tilde{\zeta} = \exp \Biggl(\frac{\sigma(1+ \cos( \theta) \lambda^2}{k_B T} \Biggr), \\
        H' = \frac{\Delta \Tilde{P} + \cos(\theta)}{H+ \frac{\Tilde{\zeta}}{R}}, \\
         H(\Delta s) = H_0 + \Delta H, \\
        \Delta t = \tau \Delta s, \\
        \quad H(0)=H_0,
         \\
         \quad  b_l \leq \theta \leq b_u,
    \end{align*}
where $H_0 >0$. \\
    
    Furthermore, the consideration of additional inertial effects is of interest. Therefore, we use the ODE \eqref{eqn:blake2004-model-nondim} instead of \eqref{eqn:governing-equation-without-inertia} and consider a dimensionless initial velocity $H'(0) \geq 0$, which is calculated by the equation \eqref{eqn:governing-equation-without-inertia}, and thus coincides with the initial velocity of the non-inertial model. Overall, we obtain the following optimization problem with inertial effects for the contact angle
     \begin{align*}
        \min_{\theta} \Delta t \quad \text{s.t.} 
        \quad \quad \quad \quad \quad  \quad \quad \quad \quad \quad \quad \quad \quad \quad  \Tilde{\zeta} = \exp \Biggl(\frac{\sigma(1+ \cos( \theta) \lambda^2}{k_B T}\Biggr), \\
       H'' = 32 \text{Oh}^2 \cdot \frac{\Delta \Tilde{P} + \cos(\theta)-HH'- \frac{\Tilde{\zeta}}{4}H' - \frac{H'^2}{32 \text{Oh}^2}}{H}, \\ 
        H(\Delta s) = H_0 + \Delta H, \\
        \Delta t = \tau \Delta s , \\
         H(0)=H_0, \quad
        H'(0)= \frac{\Delta \Tilde{P} + \cos(\theta)}{H_0 + \frac{\Tilde{\zeta}}{4}}, \\
         \quad  b_l \leq \theta \leq b_u,
    \end{align*}
where $H_0>0$ and Oh$= \frac{\eta}{\sqrt{R \rho \sigma}}>0$.

    \section{Numerical Results}

    \begin{table}
\centering
    \begin{subtable}
      \centering
    \begin{tabular}{c|c|c|c|c}
        \hline
               solution & shear viscosity $\eta$  & surface tension $\sigma$ & density $\rho$  & $\lambda$  \\
                        (mass $\%$ glycerol) & (mPa $\cdot$ s) &  (mN/m) &  (kg/m$^3$) &  (nm)   \\
              \hline
             0 & 1.0 & 72.46 & 997 & 0.62   \\
             \hline
             20 & 1.8 & 70.62 & 1043 & 0.62 \\
             \hline
             40 & 3.6 & 67.60 & 1090 & 0.62  \\
             \hline
             60 & 10.2 & 66.15 & 1140 & 0.62 \\
              \hline
             80 & 48.5 & 64.29 & 1196 & 0.599 \\
              \hline
             90 & 158.6 & 64.36 & 1225 & 0.641 \\ 
              \hline
        \end{tabular}
        \caption{Physical properties on glass for water-aqueous glycerol solutions}
        \label{table1}
        \end{subtable}
        \vspace{1.5em}
    \begin{subtable}
        \centering
        \begin{tabular}{c|c|c|c|c|c|c}
        \hline
             solution (mass $\%$ glycerol) & 0 & 20 & 40 & 60 & 80 & 90 \\
              \hline
             Ohnesorge & 0.0037 &    0.0066  &  0.0133 &    0.0371 &    0.1749 &    0.5648  \\
             \hline
        \end{tabular} 
\caption{Ohnesorge number for different water-aqueous glycerol solutions}
\label{tabelle2}
    \end{subtable}
    \end{table}

The optimization is performed in \textit{MATLAB}\textsuperscript{\textregistered} using an interior point algorithm \textit{fmincon} without providing gradient information and the ordinary differential equation solver \textit{ode45} based on a Runge-Kutta method. In the first step of the optimization, the friction coefficient $\Tilde{\zeta}$ based on the contact angle $\theta$ is computed and in the case of inertial effects, we additionally calculate the initial velocity $H'$. Then, the optimization problem is solved by computing the analytical solution or by \textit{ode45}, using adaptive time steps for the optimization of the ODE. This means that in each optimization step, $\Delta s$ is first evaluated on a coarser time mesh and use this result to compute a finer time mesh for a more accurate evaluation of the ODE.  More precisely, the new time period is chosen ten times larger than the result and the number of time steps two times larger. Further, we simplify the solving procedure for the second order ODE \eqref{eqn:blake2004-model-nondim} by substituting $\Tilde{H}=\frac{1}{2}H^2$ and obtain an ODE of the form
\begin{equation*}
    \Tilde{H}'' = 32 \text{Oh}^2 \Biggl( \Delta \Tilde{P} + \cos(\theta)-\Tilde{H}'- \frac{\Tilde{\zeta} \Tilde{H}'}{4\sqrt{2 \Tilde{H}}} \Biggr). 
\end{equation*}

The dimensionless time $\Delta s$ is then approximated by using the solution of the ODE and a linear interpolation. In the last step of the optimization, we compute $\Delta t = \tau \Delta s$.\\
As an application of the optimization, water-aqueous glycerol solutions from \cite{Duvivier2011} with different mass fractions of glycerol are considered, see Table \ref{table1}. These solutions have the property that the viscosity increases while the surface tension and density change only slightly. Furthermore, $k_B$ is the Boltzmann constant and we choose a temperature of $T=297.15$ K, a pressure difference of $\Delta P =150$ Pa, and an initial equilibrium contact angle of $\theta_0=100^{\circ}$. For the setup of the radius $R$, the initial length $H_0$, and $\Delta H$, the same values as in \cite{Blake2004} are selected. Hence, we assume a capillary radius of $R = 1$ mm, an initial length $H_0=10$, and $\Delta H = 100$ so that $h_0 =1$ cm and $\Delta h=10$ cm. \\
    \begin{figure}[ht]
	\begin{minipage}[c]{.48\linewidth}
	\includegraphics[width=\textwidth]{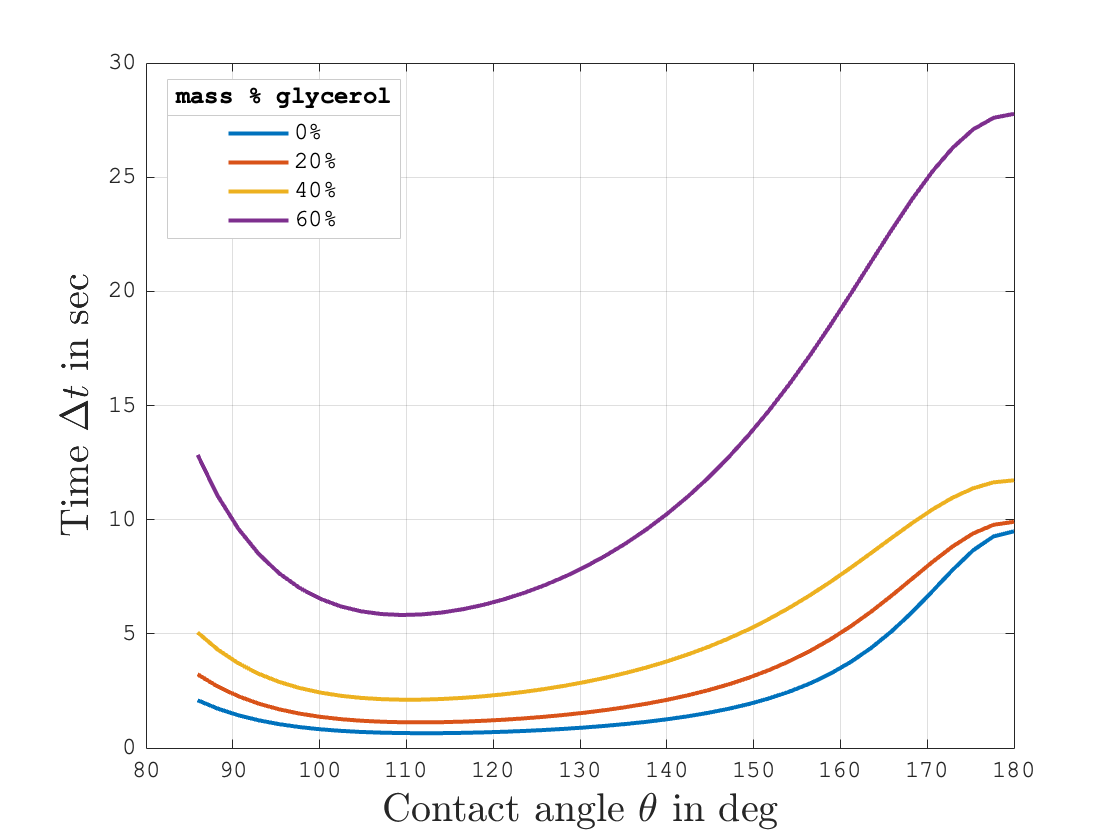}   
	\end{minipage}
	\hfill
	\begin{minipage}[c]{.48\linewidth}
	\includegraphics[width=\textwidth]{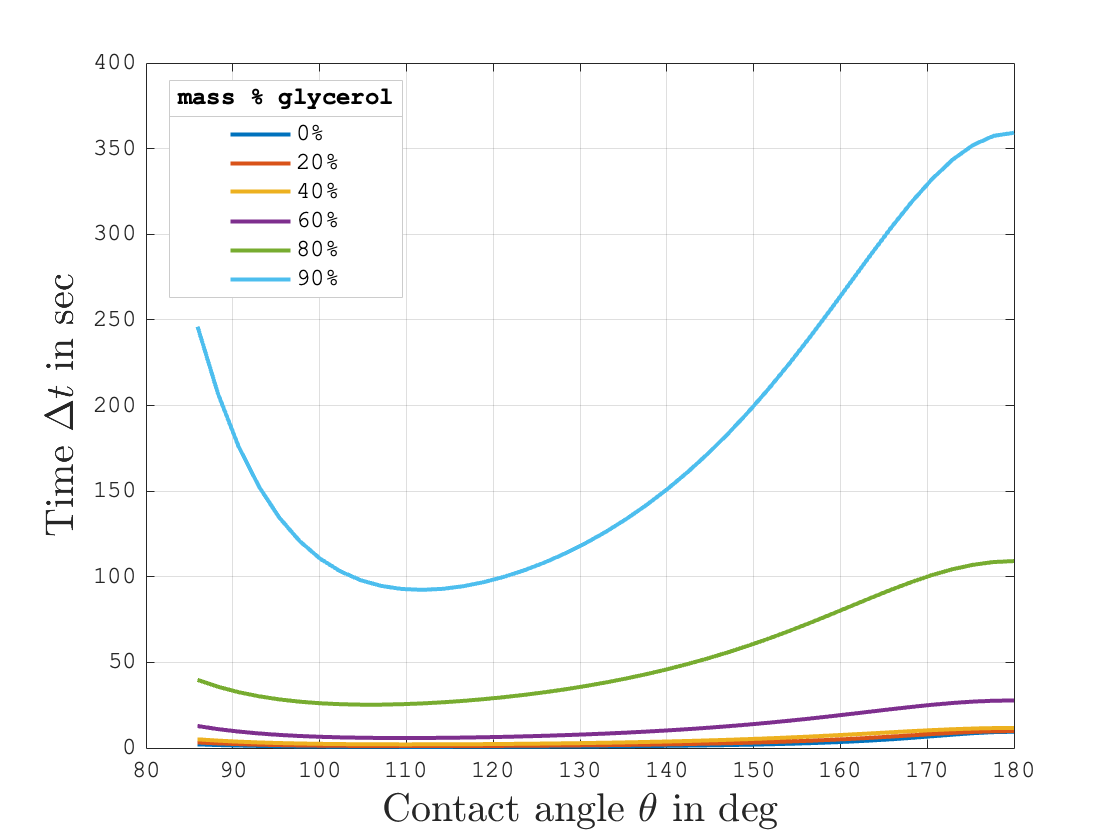}
 	\end{minipage}
  \caption{The time $\Delta t$ required for water-aqueous glycerol solutions to wet $\Delta h=10$cm depending on the contact angle $\theta$. The time $\Delta t$ is calculated by numerically solving the non-inertial model.}
	\label{fig:time_theta}
\end{figure}

The results of the optimization are presented in the Tables \ref{tabelle3}, \ref{tabelle4}, \ref{tabelle5} and \ref{tabelle6}, showing the comparison between the optimization of the analytical solution and the numerical solution of the non-inertial model as well as the numerical solution of the model including inertial effects. It can be observed that the optimal contact angle $\theta$ obtained using the analytical and numerical solution of the non-inertial model are almost identical. Hence, the minimized wetting times are also nearly the same.  Furthermore, the optimal contact angle $\theta$ of the model with inertial effects approach the analytical solution of the non-inertial model with increasing viscosity, i.e.\ with increasing glycerol mass fraction and Ohnesorge number, see Tables \ref{tabelle2}, \ref{tabelle3} and \ref{tabelle5}. \\

\begin{table}
\centering 

    \begin{subtable}
    \centering \begin{tabular}{ccccccc}
        \hline
            solution (mass $\%$ glycerol) &  $\#$ It. & $\#$ f & $\Delta t$ & $\theta_{\text{opt}}$ & first-order optimality & time \\
              \hline
               0 & 7&18  & 6.474e-1    &112.09   & 1.688e-08   &0.44    \\
               \hline
               20 & 7 & 18 & 1.122 & 111.44 & 1.672e-08 & 0.26 \\
               \hline
                40 & 8 & 20 & 2.113 & 110.29 & 1.644e-08 & 0.23 \\
                \hline
                60 & 8 & 20 & 5.823 & 109.69 & 9.925e-08 & 0.26 \\
                 \hline
                 80 & 8 & 21 & 2.529e+1 & 105.84 & 3.826e-07 & 0.20 \\
                  \hline
                   90 & 8 & 20 & 9.234e+1 & 111.70 & 1.678e-06 &0.24 \\
             \hline
        \end{tabular}
              \caption{Results of the optimization for the analytical solution of the non-inertial model.} \label{tabelle3}
    \end{subtable}
    
    \vspace{1.5em} 

    \begin{subtable}
    \centering \begin{tabular}{ccccccc}
        \hline
            solution (mass $\%$ glycerol) & $\#$ It. & $\#$ f & $\Delta t$ & $\theta_{\text{opt}}$ & first-order optimality & time \\
              \hline
              0 & 7 & 18  & 6.475e-1   &112.08   & 1.687e-8    & 1.90   \\
              \hline
             20 & 7 & 18 & 1.122 & 111.43 & 1.671e-8 & 1.58 \\
              \hline
             40 & 8 & 20 & 2.113 & 110.27 & 4.042e-8 & 1.39\\
              \hline
              60 & 9 & 25 & 5.824 & 109.67 & 5.644e-8 & 1.58 \\
             \hline
             80 & 8 & 21 & 2.529e+1 &  105.82 & 1.337e-7 & 1.26\\
             \hline
             90 & 10 & 24 & 9.235e+1 & 111.68 & 2.453e-6 & 1.89 \\
             \hline
        \end{tabular}
\caption{Results of the optimization for the numerical solution of the non-inertial model.}
\label{tabelle4}
    \end{subtable}
    \vspace{1.5em}

\begin{subtable}
    \centering \begin{tabular}{ccccccc}
        \hline
             solution (mass $\%$ glycerol) &$\#$ It. & $\#$ f & $\Delta t$ & $\theta_{\text{opt}}$ & first-order optimality & time \\
              \hline
              0& 7& 18  & 7.523e-1 & 111.08   & 1.663e-8  & 2.68   \\
              \hline
              20& 8 & 20 & 1.184 & 111.11 & 1.673e-8 & 2.57 \\
             \hline
            40& 8 & 20 & 2.146 & 110.21 & 1.642e-8 & 2.22 \\
             \hline
            60& 9 & 22 & 5.835 & 109.68 & 3.698e-8 & 3.17 \\
             \hline
             80&12 & 31 & 2.530e+1 & 105.84 & 3.448e-6 & 7.83 \\
             \hline
             90&12 & 33 & 9.234e+1 & 111.70 & 1.699e-4 & 72.16 \\
             \hline
             
        \end{tabular}
\caption{Results of the optimization including inertial effects with an initial velocity $\dot{h}>0$.}
\label{tabelle5}
    \end{subtable}
    \vspace{1.5em}

    \begin{subtable}
    \centering \begin{tabular}{ccccccc}
        \hline
              solution (mass $\%$ glycerol) & $\#$ It. & $\#$ f & $\Delta t$ & $\theta_{\text{opt}}$ & first-order optimality  & time \\
              \hline
             0& 7 & 18  & 7.524e-1 & 111.07 & 2.168e-8 & 2.73 \\
             \hline
             20& 8 & 20 & 1.184 & 111.11 & 1.673e-8 & 2.54 \\
             \hline
            40& 8 & 20 & 2.146 & 110.21 & 1.642e-8 & 2.23\\
             \hline
            60& 8 & 20 & 5.835 & 109.68 & 1.498e-7 & 3.02 \\
             \hline
            80& 13 & 33 & 2.529e+1 & 105.84 & 3.090e-7 & 9.01 \\
             \hline
             90&12 & 30 & 9.234e+1 & 111.70 & 3.355e-5 & 77.93 \\
             \hline
             
        \end{tabular}
\caption{Results of the optimization including inertial effects with an initial velocity $\dot{h}=0$.}
\label{tabelle6}
    \end{subtable}
\end{table}

The notation \grq $\#$ It.' denotes the number of iterations, \grq $\#$ f' is the number of function
evaluations, $\Delta t$ is the value of the objective function at the optimal contact
angle $\theta_{\text{opt}}$ and \textit{time} is the overall execution time
in seconds. \\

\section{Summary and Outlook}
In this note, we derived analytical solutions of the model by Blake and De~Coninck \cite{Blake2004}, i.e.
\begin{align}\label{eqn:governing-eq-conclusion}
\Delta \tilde{P} + \cos \theta_0 = H H' + \frac{1}{32 \, \text{Oh}^2} (H H')' + \frac{1}{4} \tilde{\zeta} H',
\end{align}
in the limiting cases of negligible inertia (i.e.\ $\varepsilon =1/(32 \, \text{Oh}^2)\rightarrow 0$) and negligible dimensionless contact line friction $\tilde\zeta = \zeta/\eta$. By using the model by Blake et al. for the contact line friction, i.e.\
\begin{align*}
\zeta = \eta \frac{V_l}{\lambda^3} \exp\left( \frac{\sigma(1+\cos \theta_0)\lambda^2}{k_B T} \right),
\end{align*}
one can show that a pressure dependent optimum contact angle exists, which maximizes the imbibition speed. The validity of the analytical expression \eqref{eqn:optimal_contact_angle_martic} for the optimal contact angle is confirmed via a comparison to a numerical optimization. Moreover, the effect of inertia on the optimal contact angle was investigated. As a physical example, we considered different water-aqueous glycerol solutions. As expected, it is found that the optimization results of the model with inertial effects approach the optimal values of the non-inertial model as the viscosity increases.\\
\\
Several improvements of the model \eqref{eqn:governing-eq-conclusion} may be investigated in the future. For example, for high imbibition rates, a non-linear model like the Molecular Kinetic Theory \cite{Blake1969, Blake2004, Blake2015} should be applied rather than the linear friction model \eqref{eqn:linear-cl-friction}. The analytical solution derived in this work does not cover this more general case. Moreover, it was recently shown that the hydrodynamic friction due to the viscous dissipation in the flow close to the contact line should be added to the model (see \cite{Delannoy2019,PreprintFricke2023a,Fricke2023}). This is a rather important improvement of the model but, fortunately, the resulting model has still the same mathematical structure like \eqref{eqn:governing-eq-conclusion} (at least for small contact line velocities). In this case, the analytical solution \eqref{eqn:imbibition_time_martic_type} is still applicable, but $\tilde\zeta$ will then contain contributions from both the contact line friction and the hydrodynamic friction. This case shall be studied in detail in the future.

\section*{Acknowledgements}
We acknowledge the financial support by the German Research Foundation (DFG) within the  Collaborative Research Centre 1194 (Project-ID 265191195).

\end{document}